\documentclass[fleqn,usenatbib]{mnras}

\usepackage{newtxtext,newtxmath}

\usepackage[T1]{fontenc}
\usepackage{ae,aecompl}

\usepackage{graphicx}	
\usepackage{amsmath}	
\usepackage{amssymb}	

\usepackage{esvect} 

\usepackage{xcolor}

\title[Deep learning DES mass maps]{Deep learning dark matter map reconstructions from \hspace{2cm} DES SV weak lensing data}

\author[N. Jeffrey et al.]{
Niall Jeffrey,$^{1}$\thanks{E-mail: niall.jeffrey.15@ucl.ac.uk}
Fran\c{c}ois~Lanusse$^{2}$,
Ofer~Lahav$^{1}$,
Jean-Luc~Starck$^{3}$ 
\vspace{.05cm} \\
$^{1}$ Department of Physics \& Astronomy, University College London, Gower Street, London, WC1E 6BT, UK\\
$^{2}$ Berkeley Center for Cosmological Physics and Department of Physics, University of California, Berkeley, CA 94720 \\
$^{3}$ AIM, CEA, CNRS, Universit\'{e} Paris-Saclay, Universit\'{e} Paris Diderot, Sorbonne Paris Cit\'{e}, F-91191 Gif-sur-Yvette, France \\
}

\date{Accepted XXX. Received YYY; in original form ZZZ}

\pubyear{2019}

\begin{document}
\label{firstpage}
\pagerange{\pageref{firstpage}--\pageref{lastpage}}
\maketitle

\begin{abstract}
We present the first reconstruction of dark matter maps from weak lensing observational data using deep learning. We train a convolution neural network (CNN) with a Unet based architecture on over $3.6\times10^5$ simulated data realizations with non-Gaussian shape noise and with cosmological parameters varying over a broad prior distribution. We interpret our newly created DES SV map as an approximation of the posterior mean $P(\kappa | \gamma)$ of the convergence given observed shear. Our \href{https://github.com/NiallJeffrey/DeepMass}{DeepMass}\footnotemark \ method is substantially more accurate than existing mass-mapping methods. With a validation set of 8000 simulated DES SV data realizations, compared to Wiener filtering with a fixed power spectrum, the DeepMass method improved the mean-square-error (MSE) by 11 per cent. With N-body simulated MICE mock data, we show that Wiener filtering with the optimal known power spectrum still gives a worse MSE than our generalized method with no input cosmological parameters; we show that the improvement is driven by the non-linear structures in the convergence. With higher galaxy density in future weak lensing data unveiling more non-linear scales, it is likely that deep learning will be a leading approach for mass mapping with Euclid and LSST.
\end{abstract}




\footnotetext{github.com/NiallJeffrey/DeepMass}

\section{Introduction} \label{sec:intro}

The evolving cosmological density field is rich in information about the cosmological model of the Universe, its unknown parameters, and cosmic web-dependent astrophysics. Though the largest fraction of the density is invisible dark matter, the gravitational lensing effect of galaxies can be used to infer fluctuations in the total foreground matter distribution. Accurate \textit{mass maps} will be essential for the science goals of the upcoming LSST survey and the ESA Euclid mission.

The maps considered in this paper are of the  two-dimensional convergence, $\kappa$, a weighted projection of the matter density field in the foreground of the observed galaxies. Recovering the convergence from the measured galaxy shapes, known as observed shear $\gamma_{\rm obs}$ in the weak lensing regime, is an ill-posed inverse problem, troubled by survey masks (missing data) and galaxy ``shape noise''.

A typical principled approach to reconstructing more accurate mass maps in the presence of noisy, masked shear data is to use physically motivated priors. In~\cite{jeffrey2018} it was shown that using either Gaussian priors or ``halo model'' sparsity priors for $\kappa$ improved the accuracy of the reconstructions with Dark Energy Survey Science Verification (DES SV) data. Implemented methods include using log-normal~\citep{lognormal_prior} priors or E-mode priors~\citep{Mawdsley_ff}.

However, all of these priors take functional forms that only approximate the true object of interest, the prior on the convergence field $P(\kappa | \mathcal{M})$ (with model assumptions $\mathcal{M}$). These approximations are necessary because we cannot represent the probability distribution of the non-linear density field in closed form. For example, we cannot characterise it uniquely in terms of its moments~\citep{carron_moments}. Even if the true, unapproximated prior were available, evaluation via direct calculation would likely be intractable.

Fortunately we can still draw realizations of convergence maps from the prior distribution $P(\kappa)$ in the form of simulations, which provides opportunity to a new generation of methods based on deep learning. Such an approach has been simultaneously proposed by~\cite{Shirasaki2018}, where a conditional adversarial network was used to learn a mapping from noisy convergence maps to an estimate of the noise-free convergence.

In this work, we propose a deep learning method to estimate the posterior mean of the convergence map from observed weak lensing shear measurements. In section~\ref{sec:results} we demonstrate our method on simulations and DES SV data.

\section{Weak gravitational lensing}

\subsection{Shear and convergence}

Given a distribution of source galaxies $n$ in radial comoving distance $\omega$, the convergence at position $\vv{\phi}$ on the sky is given by a weighted integral of the density

\begin{equation} \label{eq:kappa_projectd}
\kappa(\vv{\phi}) = \frac{3 H_0^2 \Omega_m}{2} \int_0^\infty  \Big[ \int_0^\omega \mathrm{d} \omega' \frac{\omega' (\omega - \omega')}{\omega} \frac{\delta(\vv{\phi}, \omega')}{a(\omega')} \Big] n(\omega) \mathrm{d} \omega \ ,
\end{equation}

\noindent where $H_0$ is the present value of the Hubble parameter, $a$ is the cosmological scale factor, $\Omega_m$ is the matter density parameter, and $\delta$ is the overdensity.

We express the linear data model in matrix notation,

\begin{equation} \label{eq:matrix}
\boldsymbol{\gamma} = \mathbf{A} \boldsymbol{\kappa} + \mathbf{n} \ ,
\end{equation}

\noindent where $\mathbf{n}$ is a vector of noise per pixel. The matrix operator acting on the convergence $\mathbf{A} \boldsymbol{\kappa}$ is the shear contribution due to lensing~\citep{bartelmann_schneider}. In this formulation, the elements $\boldsymbol{\gamma}$ are the complex shear measurements binned into angular pixels in a two-dimensional image format.

We do not take into account the second order effects of reduced shear~\citep{seitz_schneider1}, flexion~\citep{bacon_flexion} or intrinsic alignments~\citep{kirk_ia}. However, the deep learning approach taken in this paper is extremely flexible; as long as an effect can be modelled and included in the training data, it will be taken into account in the mass map reconstruction. This is not generally true of other methods. For example, flexion requires reformulations of methods (e.g. \citealt{glimpse2016}). Additionally, noise per pixel is invariably approximated as Gaussian, which we do not assume in our deep learning approach.

\subsection{Previous mapping approaches}

The original mass mapping approach by~\cite{ks93} was a direct deconvolution. In practice Kaiser-Squires (KS) inverts the matrix $\mathbf{A}$ in Fourier space, where the matrix is diagonal. As this deconvolution is across a finite space, the edges of the data and internal masks introduce artefacts. KS is further troubled by the noise term in equation~\ref{eq:matrix}, which it does not take into account. 

In a Bayesian framework we may wish to consider the posterior distribution of the convergence $\boldsymbol{\kappa}$ conditional on the observed shear $\boldsymbol{\gamma}$

\begin{equation}
    P(\boldsymbol{\kappa} | \boldsymbol{\gamma}, \mathcal{M}) = \frac{P(\boldsymbol{\gamma} | \boldsymbol{\kappa}, \mathcal{M}) \ P(\boldsymbol{\kappa} | \mathcal{M})}{P(\boldsymbol{\gamma} | \mathcal{M})} \ \ ,
\end{equation}

\noindent The denominator $P(\boldsymbol{\gamma})$ is a Bayesian evidence term conditional on model $\mathcal{M}$. The first factor of the numerator is the likelihood $P(\boldsymbol{\gamma} | \boldsymbol{\kappa}, \mathcal{M})$, which encodes our noise model. The second term is the prior $P(\boldsymbol{\kappa} | \mathcal{M})$, a possible selection of which was discussed in section~\ref{sec:intro}. 

If we believe a realization of the convergence $\boldsymbol{\kappa}$ is a realization of Gaussian random field, then the form of $P(\boldsymbol{\kappa})$ would be Gaussian. If the noise per pixel is Gaussian then the likelihood is also Gaussian, which results in a posterior distribution with both the mean and maximum given by the Wiener filter:

\begin{equation} \label{eq:wiener}
\hat{\boldsymbol{\kappa}}_{\rm w} = \mathbf{W} \boldsymbol{\gamma} = \mathbf{S}_\kappa \mathbf{A}^\dagger \big[ \mathbf{A} \mathbf{S}_\kappa \mathbf{A}^\dagger + \mathbf{N} \big]^{-1}\  \boldsymbol{\gamma}  \ ,
\end{equation}

\noindent where $\mathbf{S}_\kappa=\langle \boldsymbol{\kappa} \boldsymbol{\kappa}^\dagger \rangle$ and $\mathbf{N}=\langle \mathbf{n} \mathbf{n}^\dagger \rangle$ are the signal and noise covariance matrices respectively (\citealt{wiener1949extrapolation},~\citealt{zaroubi_wiener},~\citealt{jeffrey_heavens_fortio}). The signal covariance in harmonic space is diagonal for isotropic fields. On the sphere, its elements are given by the $\boldsymbol{\kappa}$ power spectrum, $C_{\kappa}(\ell)$. 

This Gaussian distribution is only approximately true for large scales where Gaussianity persists from the early Universe. On smaller scales, non-Gaussianity grows due to non-linear structure formation, which results in the cosmic web of the late Universe.

\begin{figure}
\hspace*{-0.5cm}
\includegraphics[width=0.513\textwidth]{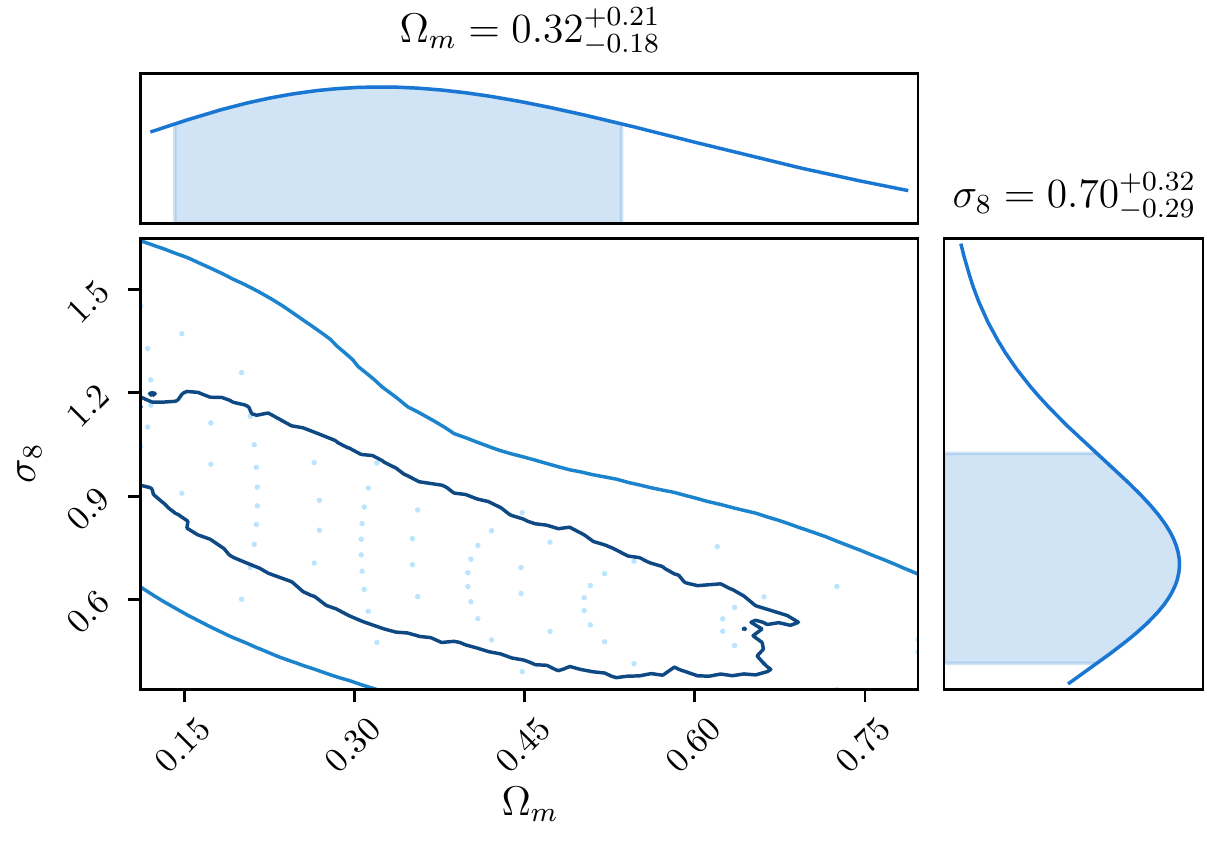}
\caption{\label{fig:params} Prior range of cosmological parameters $\Omega_m$ and $\sigma_8$ of the training data. Simulations were run at the marked points.}
\end{figure}

\begin{figure*}
\centering
\includegraphics[width=0.99\textwidth]{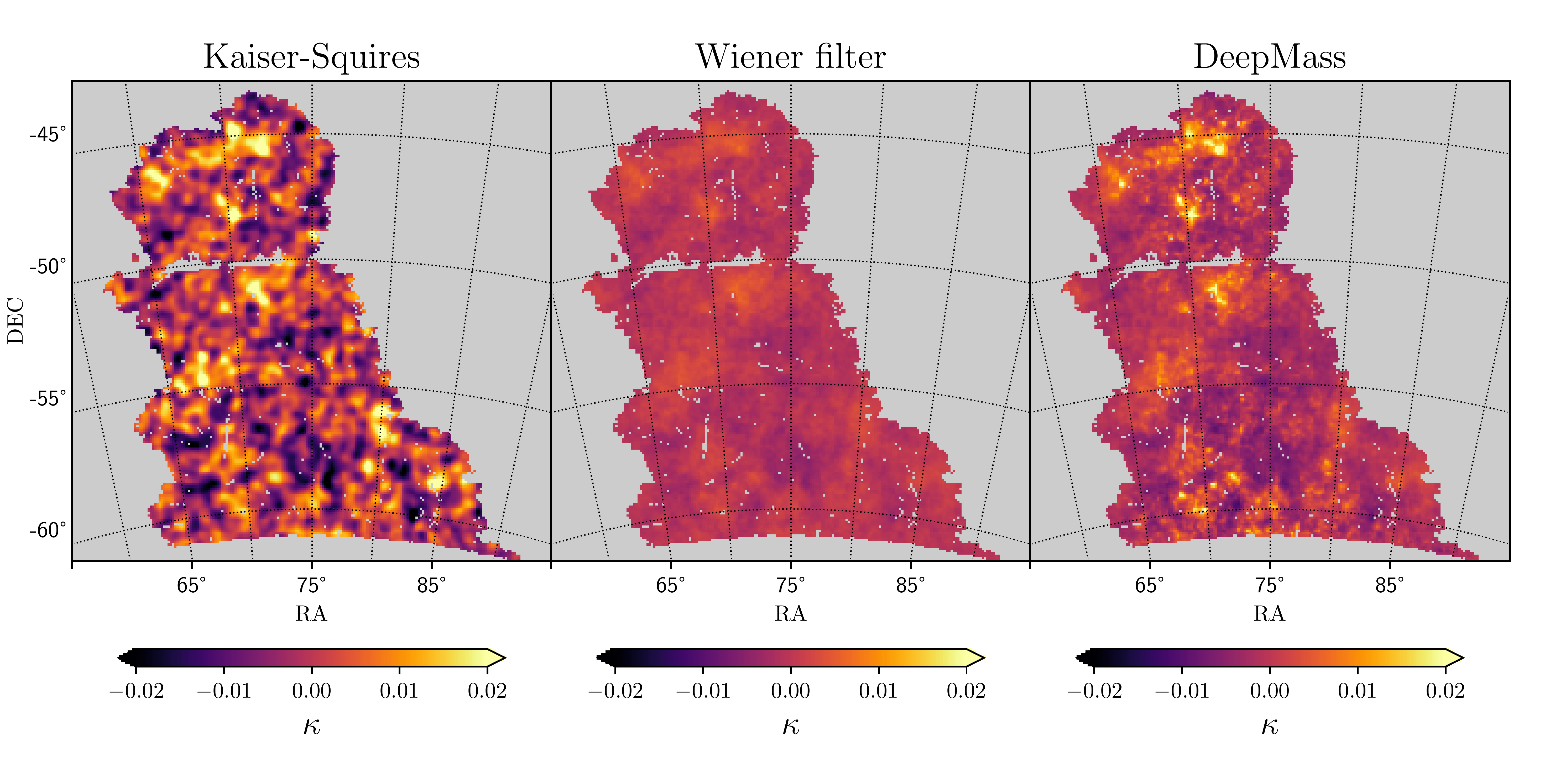}
\caption{\label{fig:des_sv_map} Convergence $\boldsymbol{\kappa}$ reconstruction from DES SV observational data with: KS, Wiener filtering, and DeepMass.}
\end{figure*} 

\section{Deep learning maps}

\subsection{Convolution neural networks}

We take a standard deep learning approach. We seek an approximation $\mathcal{F}_\Theta$ to the function that maps the pixelized shear to the convergence map

\begin{equation} \label{eq:function}
\hat{\boldsymbol{\kappa}} =  \mathcal{F}_\Theta (\boldsymbol{\gamma}) \ \ ,
\end{equation}

\noindent where the parameters of the function $\Theta$ are to be learned~\citep{deep_learning}. We learn these parameters by minimising a mean-square-error (MSE) cost function 

\begin{equation} \label{eq:loss}
J(\Theta) = || \mathcal{F}_\Theta (\gamma) - \kappa_{\rm true}||_2^2 \ \ ,
\end{equation}

\noindent evaluated on a set of training data which consists of pairs of realistic shear and ``truth'' (noise-free) convergence maps. If the training data ``truth'' maps are drawn from a prior distribution $P(\kappa)$, and the corresponding noisy shear map is drawn from the likelihood $P(\gamma | \kappa)$, this MSE cost function corresponds to $\mathcal{F}_\Theta (\gamma)$ being a mean\footnote{The mean posterior is not generally the {\it maximum a posteriori}}  posterior estimate~\citep{jaynes07}, such that $\hat{\boldsymbol{\kappa}}$ is approximating:

\begin{equation} \label{eq:mse_mean}
\hat{\boldsymbol{\kappa}} =  \mathcal{F}_\Theta (\boldsymbol{\gamma}) = \int \boldsymbol{\kappa} \ P(\boldsymbol{\kappa} | \boldsymbol{\gamma}) \ {\rm d} \boldsymbol{\kappa} \ \ .
\end{equation}

\noindent We use a deep convolution neural network (CNN) to approximate the function $\mathcal{F}_\Theta$, where the parameters $\Theta$ are primarily elements of learned filters in convolutional layers. CNNs are particularly suited for two-dimensional image or one-dimensional time series data with translation invariant features in the underlying signal.

The CNN is a series of iteratively computed layers. At a given layer $j$ the signal $\mathbf{x}_j$ is computed from the previous layer 

\begin{equation}
\mathbf{x}_j = \rho \mathbf{M}_j \mathbf{x}_{j-1}
\end{equation}

\noindent with linear operator (e.g. convolution) $\mathbf{M}_j$ and nonlinear {\it activation} function $\rho$  (\citealt{lecun1990handwritten},~\citealt{mallat2016understanding}). The output of a layer is sometimes called a {\it feature map}. 

Due to their additional layers, deep architectures are often able to learn features with greater complexity than shallow architectures and therefore can better approximate the target function. For a general overview of deep learning and neural networks we recommend~\cite{deep_learning}. 

\begin{figure*}
\centering
\includegraphics[width=0.99\textwidth]{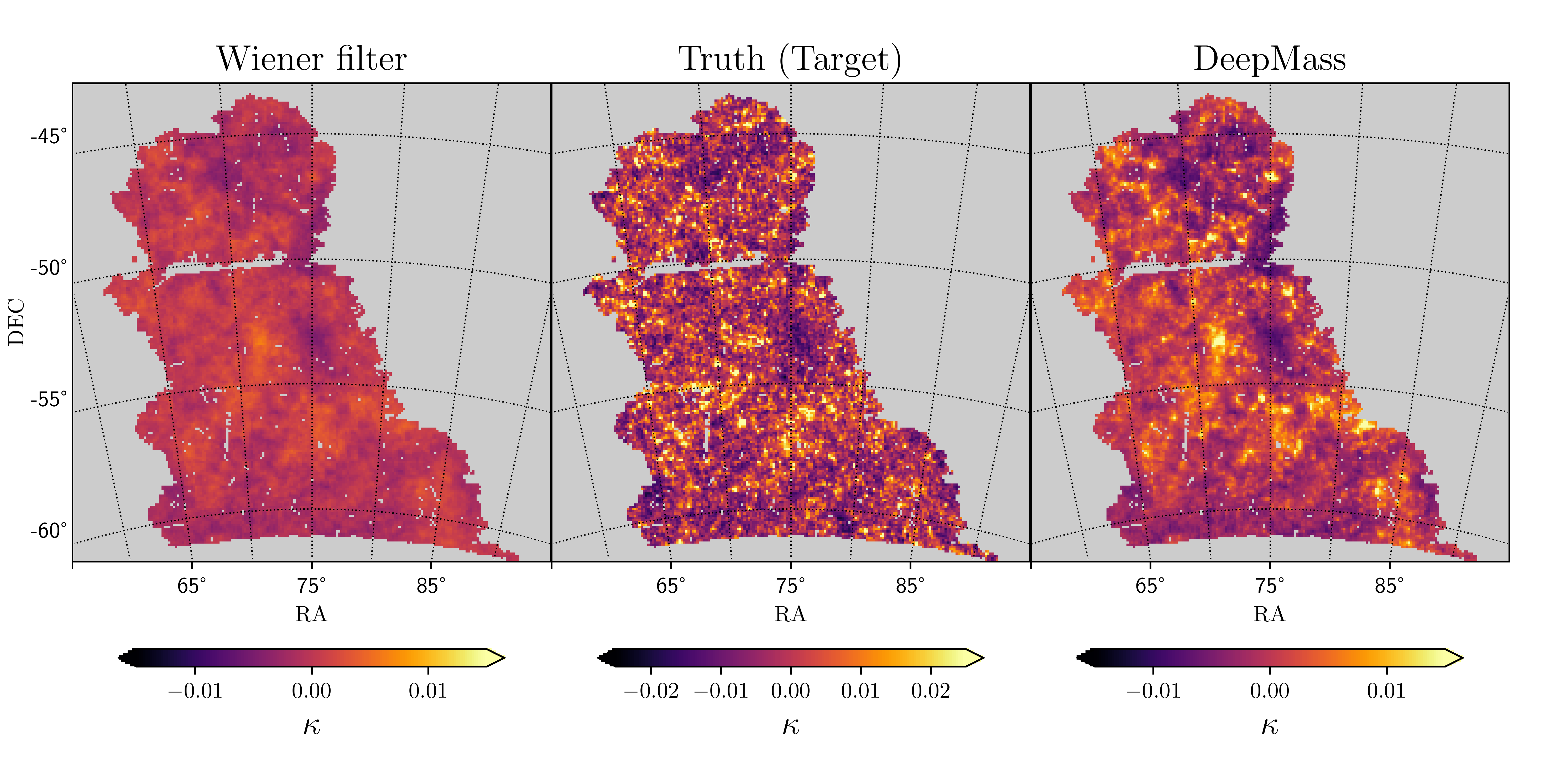}
\caption{
{\label{fig:sims_map} Example {\sc l-picola} validation simulation ({\it centre}) and the corresponding Wiener ({\it left}) and DeepMass ({\it right}) reconstructions. The colour scale for the truth (target) is larger to accommodate the larger dynamic range and make comparison easier by eye.}}
\end{figure*}

\subsection{DeepMass architecture} \label{sec:architecture}

Our DeepMass architecture is based on the Unet~\citep{ronneberger2015u}, which has a so-called \textit{expanding path} and \textit{contracting path}. The DeepMass contracting path differs from the original Unet: usually convolutions and activation are followed by a \textit{max pooling} operation to downsample the images, whereas we use \textit{average pooling}~\citep{geron2017hands}. With each downsampling operation, the images decrease in resolution, but the 3$\times$3 filters cover more angular size of the image. The convolution after a pooling operation therefore has a receptive field that covers larger physical features in the convergence $\boldsymbol{\kappa}$ map. 
 
There are similarities between Unet architectures and sparse recovery methods. These consider representations where the solution is sparse and employ transforms which are fixed (e.g. Fourier, wavelets) or 
learned from data, and optimization is solved using proximal theory~\citep{starck:book2015}. The Unet expanding and contracting path are very similar to synthesis and analysis concepts in sparse representations. This has motivated the use of wavelets to implement the Unet average pooling and the expanding path \citep{ye2018deep,han2018framing}. There are nevertheless significant differences: Unets can learn rich sets of features (corresponding to sparse dictionaries) from large training data sets, and the CNN implementation of non-linearity. 

We differ from the original Unet by not using padding in the convolutional layers, as the edge of our data mask is already many pixels away from the edge of the square image. This choice means that output of a convolution has the same image dimensions as the input.

The full architecture and code can be seen online: \href{https://github.com/NiallJeffrey/DeepMass}{DeepMass$^\dagger$}. We have added \textit{Batch Normalization} layers~\citep{ioffe2015batch} after each convolutional layer; without this, training often became stuck in local minima of the cost function with respect to the parameters $\Theta$. For all layers, except for the final, we use the rectified linear unit (ReLU) activation. In the final layer we use a sigmoid function, which forces the output to be between 0 and 1 (inputs and outputs are correspondingly rescaled).

For simplicity and memory efficiency, we aimed to work with real (32-bit) numbers, thus necessitating an initial operation acting on the complex shear $\boldsymbol{\gamma}$. The best results came from using a fixed Wiener filter operation before the first convolution (rather than KS, as might be expected). This is equivalent to the first layer having $\mathbf{M}_{j=0} = \mathbf{W}$ and $\rho = 1$, with no free parameters. We could also interpret the Unet after the initial Wiener operation as $ \mathcal{G}_\Theta$ where $\mathcal{F}_\Theta (\gamma)  =  \mathcal{G}_\Theta ( W(\gamma))$.  The Wiener filter used a power spectrum with cosmological parameters $\sigma_8$ and $\Omega_m$ fixed at the mean of the marginal posterior distributions from DES Y1 analysis~\citep{des_year1}. The flat sky power spectrum was an average of 102 power spectra of projected patches.

\subsection{Training data} \label{sec:training}

\subsubsection{{\sc l-picola} simulations}
The training data is derived from 74 independent dark matter simulations, with each simulation covering an octant of the sky. The simulations used a standard flat $\Lambda$CDM cosmological model with $H_0=70\ {\mathrm{km \ Mpc^{-1} s^{-1}}}$. The scalar spectral index and baryon density were fixed at $n_s=0.95$ and $\Omega_b =0.044$ respectively. The values of $\Omega_m$ and the amplitude parameter $\sigma_8$ are distributed on a non-Euclidean grid with distances between points giving a density according to our prior $P(\sigma_8$, $\Omega_m)$ as shown in figure~\ref{fig:params}. Weak lensing constraints are most sensitive to combinations of this pair of parameters, so we avoid overfitting to a single cosmology by varying them in the training data.

To generate a convergence map from a simulation, the matter particles were binned using the {\sc healp}ix~\citep{healpix} pixelisation of the sphere with {\sc nside}=2048 in comoving radial shells of $50\ {\rm Mpc/h}$. The density $\rho$ map in a given redshift was converted into an overdensity $\delta = \rho / \Bar{\rho} -1$ using the average density in the shell $\Bar{\rho}$. The convergence was calculated per pixel using equation~\ref{eq:kappa_projectd}. We wish to have the $n(z)$ in the lensing kernel match the DES SV data (section~\ref{sec:results_des}), which we approximate by summing the individual posterior redshift distributions per galaxy from the BPZ photometric redshift code \citep{bpz2}. The convergence maps were downgraded to {\sc nside}=1024.

The dark matter simulations are generated using the {\sc l-picola} code~\citep{picola}, which is based on the {\sc cola}~\citep{cola} algorithm. This uses a combination of second-order Lagrangian perturbation theory (2LPT) and a Particle-Mesh (PM) which requires fewer time steps than ``full'' N-body (e.g. Gadget~\citealt{gadget}) and therefore can generate simulations more quickly. This allows more training data to be generated in a given amount of compute time.

We used a 1250 Mpc$/h$ comoving simulation box, $768^3$ particles, and a $1536^3$ grid. A $z<1.6$ lightcone was generated with up to four box replicates, using 30 time steps from $z=20$. The initial conditions used~\citealt{eisenstein_hu} for the linear matter power spectrum.

The drawback of this approach is the accuracy of the dark matter distribution. The finite spatial resolution and fewer timesteps used by the {\sc cola} method particularly affects small distance scales. Our experiments have shown a suppression of the {\sc l-picola} power spectrum at scales of $\ell > 700$ of order 10 per cent (relative to {\sc nicaea}\footnote{nicaea.readthedocs.io}~\citep{nicaea} theory), as is expected with {\sc cola} methods. We correct the power of the {\sc l-picola} convergence by estimating the smooth part of the $C_\kappa (\ell)$ using a polynomial order 1 Savgol filter with window size 91 for each convergence map and reweighting spherical harmonics. Using the ratio of {\sc nicaea} and only the smooth part of the measured simulation power spectrum ensures that the natural fluctuations inherent in $C_\kappa (\ell)$ for a given realization are preserved.
 
\begin{figure*}
\centering
\includegraphics[width=0.99\textwidth]{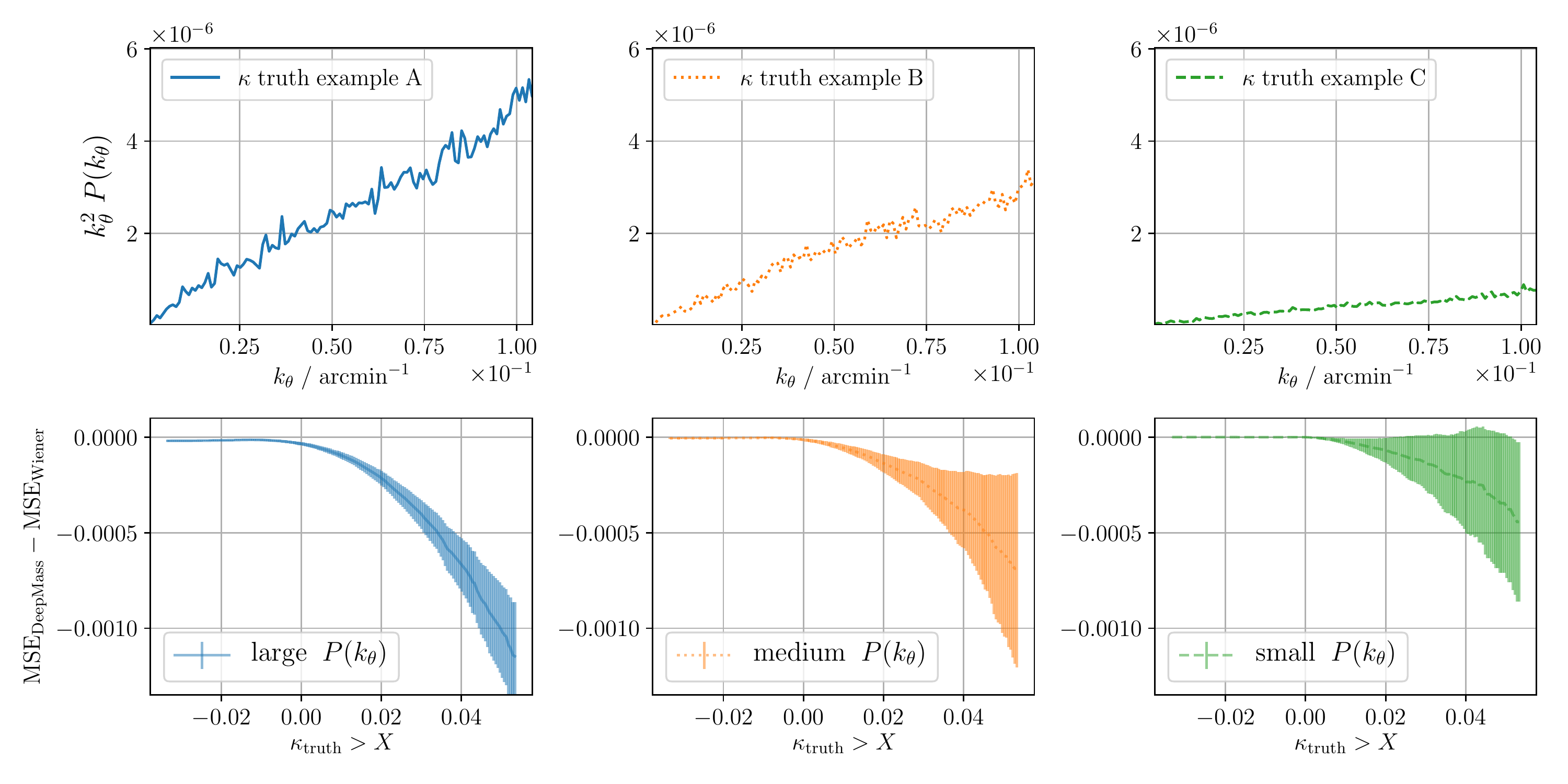}
\caption{\label{fig:kappa_cut} { The {\textit{top row}} shows the power spectra of three example convergence $\boldsymbol{\kappa}$ truth maps (A, B and C) from the validation sample. The {\textit{bottom row}} shows the corresponding change in MSE from Wiener filtering to DeepMass as a function of $\kappa$ threshold of the pixels. This shows (1) DeepMass improves over Wiener mostly in the high $\kappa$ regime, and (2) DeepMass improves over Wiener more when the underlying map has more structure (higher power).}}
\end{figure*} 

\subsubsection{Training images}

From the 74 independent {\sc healp}ix convergence maps over an octant of the sky, we generate 376,684 DES SV mock data realizations. A given realization is generated from the {\sc healp}ix convergence map by randomly choosing a position on the sphere, applying a uniform random rotation between 0 and 360 deg, and extracting a square patch using a gnomonic projection with $256^2$ pixels of size $4.5^2\ \mathrm{arcmin}^2$. If the generated image has pixels outside the octant, it is rejected. The rotation step is not to make the reconstruction rotation invariant, which happens naturally as $P(\boldsymbol{\kappa})$ is isotropic by the cosmological principle, but it is to augment the training data and learn $\mathcal{F}_\Theta$ better.

From the projected square $\boldsymbol{\kappa}$ convergence map, the complex noise-free shear map is generated using the $\mathbf{A}$ matrix from equation~\ref{eq:matrix}. The mask is applied and a random shape noise map is added. The noise map is generated by randomly shuffling the positions of galaxies in the original catalogue; this keeps the density of galaxies the same, but destroys the coherent lensing signal. This way we forward model the non-Gaussian noise inherent in the data (something that other methods do not do).

\section{Results} ~\label{sec:results}

\subsection{Dark Energy Survey SV data}\label{sec:results_des}

DES is a ground-based photometric galaxy survey, observing in the southern sky from the 4m Blanco telescope in Chile with five photometric filters~\citep{decam}. The SV (A1) data\footnote{http://des.ncsa.illinois.edu} come from an initial run of 139 deg$^2$, but with depth approximately that of the full 6 year survey~\citep{chang_sv_map}. We make a redshift cut of $0.6<z_{\rm mean}<1.2$, where $z_{\rm mean}$ is the mean of the $z$ posterior for each galaxy. Data selection choices match \cite{jeffrey2018}, although some maps appear different due to changes in pixel size and flat-sky projection.

In figure~\ref{fig:des_sv_map} we apply KS, Wiener filtering, and the trained DeepMass CNN. Kaiser-Squires uses a 10 arcmin Gaussian smoothing as in \cite{jeffrey2018}. The Wiener filtering uses a power spectrum with $\Omega_m$ and $\sigma_8$ at the mean of their respective marginal posterior distributions from the Year 1 DES cosmology result~\citep{des_year1}. The DeepMass CNN was trained using the Adam optimizer~\citep{kingma2014adam} with a learning rate = $1 \times 10^{-5}$ for 20 epochs (retraining over the full training set). The final Wiener and DeepMass maps were smoothed with a Gaussian kernel of $\sigma = 2.25$ arc min (half pixel size) to remove very small scale artefacts arising from the {\sc healp}ix projection. 

The DeepMass reconstruction clearly shows more non-linear structure than the Wiener filter. Individual peaks, which are suppressed by Wiener filtering, are resolved by DeepMass. { The accurate recovery of non-linear and peak structures using DeepMass is studied quantitatively in section~\ref{sec:results_validation}.}

\subsection{Validation on simulations} \label{sec:results_validation}

Of the originally generated training images (section~\ref{sec:training}), 8000 were reserved for validation and not used for training. One such example can be seen in figure~\ref{fig:sims_map}, with the corresponding Wiener filter and DeepMass reconstructions. As with the reconstruction from observational data,  DeepMass can be seen to recover the non-linear (cosmic-web) structure better than Wiener filtering. Compared to Wiener filtering, the MSE over all 8000 maps is improved using DeepMass by 11 per cent.

{

As DeepMass is estimating the mean of a posterior probability distribution (equation~\ref{eq:mse_mean}) there are inevitably structures that appear in the map (figure~\ref{fig:sims_map}) but do not appear in the truth and structures that appear in the truth but not in the reconstruction. Exploring the full posterior distribution, and thereby quantifying uncertainty, is a rich topic for future work.

To understand how DeepMass performs over different regimes, we can measure the MSE for a subset of pixels in the map with true value greater than some threshold, $\kappa > X$. Figure~\ref{fig:kappa_cut} shows this for three different examples (labelled A, B and C), chosen from the validation sample for their different power spectra.

The \textit{bottom row} of figure~\ref{fig:kappa_cut} shows the MSE difference between DeepMass and Wiener filtering ($\mathrm{MSE}_\mathrm{DeepMass} - \mathrm{MSE}_\mathrm{Wiener}$) for these three examples. In each case, the DeepMass improvement over Wiener filtering is shown to be driven by pixels with large true convergence $\kappa_\mathrm{truth}$. That is, DeepMass improves over Wiener filtering more for pixels with larger $\kappa$ values (with increased improvement for those with $\kappa_{\mathrm{truth}} > 0$). Compared to Wiener filtering, DeepMass is able to better reconstruct high $\kappa$ value, non-linear structures in the convergence $\kappa$ fields.

The error bars in the \textit{bottom row} of figure~\ref{fig:kappa_cut} are given by $\sqrt{\sigma_{\mathrm{MSE,\ DeepMass}}^2 + \sigma_{\mathrm{MSE,\ Wiener}}^2 ) / N_{\kappa>X}}$, where $N_{\kappa>X}$ is the number of pixels in the sample above the threshold, and $\sigma_{\mathrm{MSE}}^2$ is the variance of the measured MSE over the sample for a given method.

The three map examples in figure~\ref{fig:kappa_cut} were chosen to demonstrate performance when the underlying true maps have different power spectra (\textit{top row}). The map with the smallest power (example C) has the least improvement of DeepMass over Wiener filtering (though the improvement is still larger for high valued pixels). Example A, with a larger power spectrum, shows a much more significant improvement in MSE.

Figure~\ref{fig:kappa_cut} demonstrates that: (1) DeepMass improves over Wiener more for pixels with larger $\kappa$ value, and (2) DeepMass improves over Wiener more when the underlying map has more structure (higher power).
}

In~\cite{jeffrey2018}, use of a ``halo-model'' sparsity prior did not outperform Wiener filtering in terms of MSE, and so we expect DeepMass MSE to outperform \textsc{Glimpse}. However, MSE minimisation relates just to the posterior mean, so alternative metrics (e.g. constraints from peak statistics) remain to be explored.

Using 18 non-overlapping mock DES SV data from the MICE~\citep{mice3} simulations we apply a Wiener filter with an optimal power spectrum calculated using the known cosmological parameters (not available in real data applications). Nevertheless, without using the known cosmological parameters as input, DeepMass still recovers maps with an average of 2 per cent better MSE.

\begin{figure}
\hspace*{-0.18cm}
\includegraphics[width=0.49\textwidth]{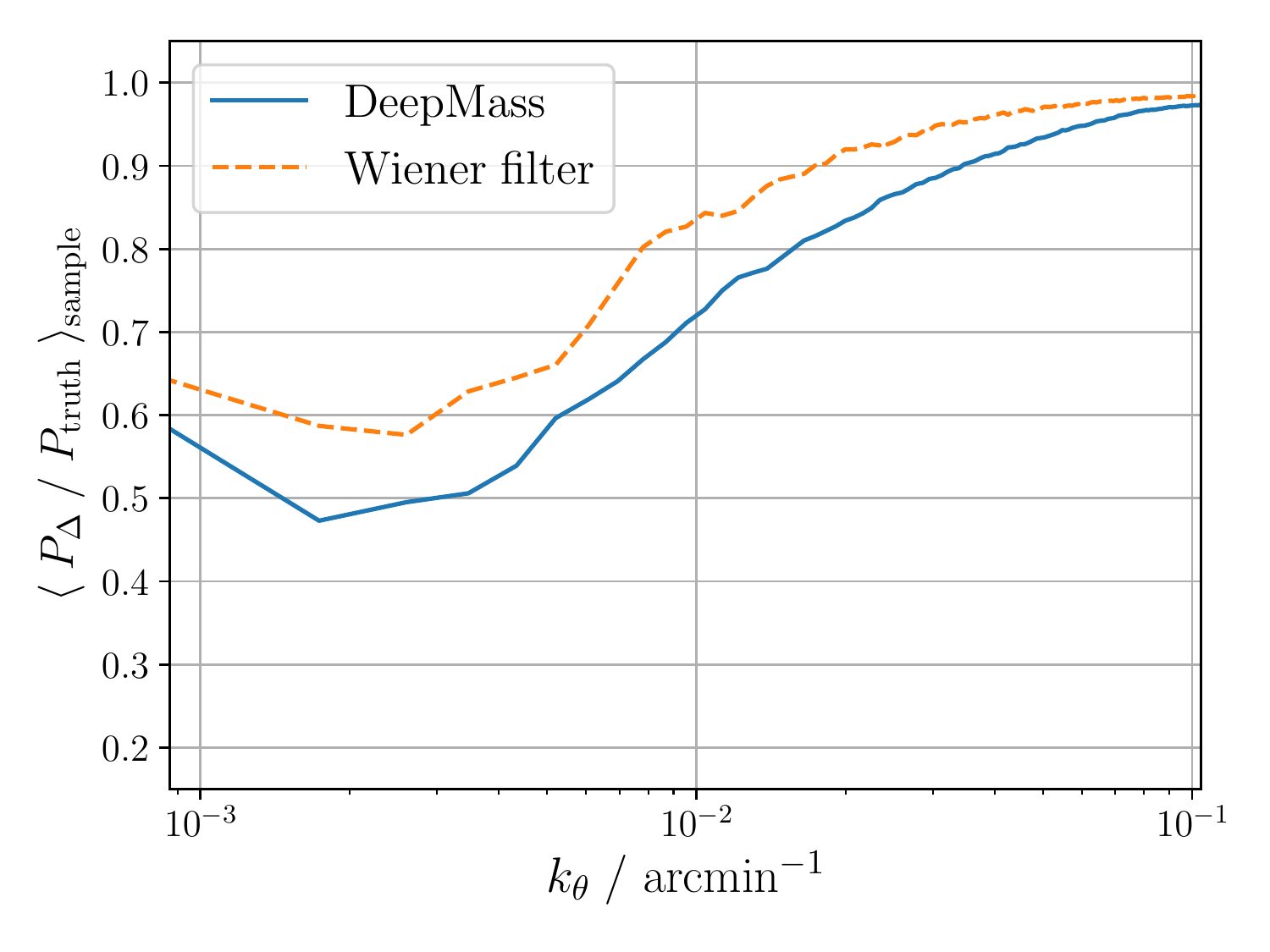}
\caption{\label{fig:residual_power} {Power spectrum of the residuals ($\boldsymbol{\kappa}_\Delta = \hat{\boldsymbol{\kappa}} - \boldsymbol{\kappa}_{\mathrm{truth}}$) normalized by the power spectrum of the truth $P_{\textrm{truth}}$. The ratio $P_\Delta /P_{\textrm{truth}}$ is averaged over 8000 maps in the validation set with cosmological parameters drawn from the prior (figure~\ref{fig:params}).}}
\end{figure}

{

The smaller improvement of DeepMass over Wiener with the MICE simulated data can be explained both by the fact that we have used the known cosmological parameters in the Wiener filter (whereas DeepMass uses no specified input cosmological parameters) and by the intrinsically low power of the MICE simulations. The MICE cosmological parameters (inc. $\Omega_m = 0.25$, $\sigma_8=0.8$) lead to a relatively low amplitude power spectrum, so DeepMass is in the regime demonstrated by example C in figure~\ref{fig:kappa_cut}. The low power effectively means that there are fewer non-Gaussian structures above a detectable signal-to-noise level. The largest improvement over Wiener filtering comes when there are more non-linear (non-Gaussian) structures.

With the same MICE simulations, and restricting ourselves to pixels where the truth is greater than two standard deviations from the mean $\kappa > 2 \sigma$ (where $\sigma$ is the standard deviation of pixels in the true map), compared to Wiener filtering, DeepMass improves the MSE by 8 per cent. As is to be expected, therefore, DeepMass improves over Wiener filtering due to its ability to reconstruct the non-linear structures in the cosmological signal.  

Figure~\ref{fig:residual_power} shows the power spectrum $P_\Delta$ of the residual maps $\hat{\kappa} - \kappa_{\mathrm{truth}}$, normalized by dividing by the true power $P_{\mathrm{truth}}$, averaged across all the maps in the validation sample. The residual power shows no particular scale at which DeepMass performs better or worse than the Wiener filter; DeepMass outperforms Wiener filtering at all length scales. 

At the smallest scales, the average $P_\Delta/P_{\mathrm{truth}}$ for both methods tend towards one. This is evidence that both reconstruction methods are suppressing structure on the smallest scales, and the residual power is tending towards the power of the true map. On these smallest scales the signal-to-noise is so low that the minimum variance reconstructions damp fluctuations.

The Wiener filter uses the same fiducial power spectrum for each map included in the averaged result shown in figure~\ref{fig:residual_power}. Although DeepMass has no input cosmology, it outperforms Wiener filtering on large scales which we may expect to be more approximately Gaussian. We can interpret this as DeepMass inferring power spectrum or cosmological parameter information from the data, information which is being used in the reconstruction.

In this work, we drew the cosmological parameters of the training data from broad priors (see figure~\ref{fig:params}) to generalize the DeepMass method for the realistic situation where the true underlying parameters are unknown. However, in future work it would be interesting to compare Wiener filtering with DeepMass trained at a single fiducial cosmology, to find whether DeepMass would outperform Wiener filtering only at small (presumably more non-Gaussian) scales. 
}

\section{Conclusion}

With DeepMass, we have presented a deep learning method to reconstruct convergence $\boldsymbol{\kappa}$ maps from shear measurements. With DES SV, we have shown the mass map reconstruction with deep learning from observational data.

By training with simulations over a broad prior distribution of cosmological parameters, we have a generalized method which needs no input cosmological parameters. This method has shown substantial improvement over Wiener filtering both qualitatively (by eye) and quantitatively (11 per cent MSE reduction on the validation data). The flexible approach also takes into account non-Gaussian noise in the weak lensing data. As our simulated training data are samples drawn from the prior $P(\kappa)$, the approach has a principled Bayesian interpretation, without the need for evaluation of closed-form priors. 

The quality of the reconstruction with these initial experiments, and its flexibility, makes the deep learning approach a preeminent candidate for mass mapping with future weak lensing surveys.

\section*{Acknowledgements}
We thank Edd Edmondson (GPU-wrangler), Ben Wandelt for a useful discussion, and Lorne Whiteway for comments. NJ and OL acknowledge STFC Grant
ST/R000476/1.




\bibliographystyle{mnras}
\bibliography{bibliog} 





\label{lastpage}
\end{document}